\documentclass[letter,structabstract]{aa} 
\usepackage{enumitem}
\usepackage{lscape,longtable}
\usepackage{graphicx}
\newcommand{\lsimeq}{{_<\atop^{\sim}}}
\newcommand{\gsimeq}{{_>\atop^{\sim}}}
\usepackage{txfonts}
\begin{document}
%
\title{PEP: first {\it Herschel} probe of dusty galaxy evolution up to $z$$\sim$3
\thanks{{\it Herschel} is an ESA space observatory with science instruments provided by European-led Principal 
Investigator consortia and with important participation from NASA.}}


   \author{C. Gruppioni\inst{1} \thanks{\email{carlotta.gruppioni@oabo.inaf.it}}
, F. Pozzi\inst{2}, P. Andreani\inst{3,4}, G. Rodighiero\inst{5}, A. Cimatti\inst{2},           
        B. Altieri\inst{6}, H. Aussel\inst{7}, S. Berta\inst{8}, A. Bongiovanni\inst{9,10}, D. Brisbin\inst{11}, A. Cava\inst{9,10}, J. Cepa\inst{9,10}, E. Daddi\inst{7}, H. Dominguez-Sanchez\inst{1}, D. Elbaz\inst{7}, 
        N. F\"orster Schreiber\inst{8}, R. Genzel\inst{8}, E. Le Floc'h\inst{7}, D. Lutz\inst{8}, G. Magdis\inst{7}, M. Magliocchetti\inst{12}, B. Magnelli\inst{8}, R. Maiolino\inst{13}, R. Nordon\inst{8},
         A.M. Per\'ez-Garc\'ia\inst{9,10}, A. Poglitsch\inst{8},  P. Popesso\inst{8}, L. Riguccini\inst{7}, 
        A. Saintonge\inst{8}, M. Sanchez-Portal\inst{6}, P. Santini\inst{13}, L. Shao\inst{8}, E. Sturm\inst{8}, L. Tacconi\inst{8}, I. Valtchanov\inst{6}      
          }
          
\institute{\centering \vskip -10pt \small \it (See online Appendix \ref{sect:affiliations} for author affiliations) }

\authorrunning{C. Gruppioni et al.}
\titlerunning{PEP FIR luminosity function}

   \date{Received March 31, 2010; accepted ???}

 
  \abstract
   {} 
   {We exploit the deepest existing far-infrared (FIR) data obtained so far by Herschel at 100 and 160 $\mu$m in the GOODS-N, as part of the PACS Evolutionary Probe (PEP) survey, to derive for the first time the evolution of the rest-frame 60-$\mu$m, 90-$\mu$m, and total IR luminosity functions (LFs) of galaxies and AGNs from $z=0$ to unprecedented high redshifts ($z$$\sim$2--3).}
   {The PEP LFs were computed using the 1/V$_{max}$ method. The FIR sources were classified by means of a detailed broad-band SED-fitting analysis and spectral characterisation. Based on the best-fit model results, k-correction and total IR (8--1000 $\mu$m) luminosity were obtained for each source. LFs (monochromatic and total) were then derived for various IR populations separately in different redshift bins and compared to backward evolution model predictions.}
   {We detect strong evolution in the LF to at least $z$$\sim$2. Objects with SEDs similar to local spiral galaxies are the major contributors to the star formation density (SFD) at $z$$\lsimeq$0.3, then, as redshift increases, moderate SF galaxies -- most likely containing a low-luminosity AGN -- start dominating up to $z$$\simeq$1.5. At $z$$\gsimeq$1.5 the SFD is dominated by the contributions of starburst galaxies. In agreement with previous findings, the comoving IR LD derived from our data evolves approximately as $(1+z)^{3.8\pm�0.3}$ up to $z$$\sim$1, there being some evidence of flattening up to $z$$\sim$2.}
   {}


   \maketitle
%

\section{Introduction}
Extragalactic surveys in the infrared (IR) represent a key tool for understanding galaxy formation and evolution, the luminosity function (LF) computed at different redshifts being the most direct method for exploring the evolution of a galaxy population. Deep surveys with {\em IRAS} and {\em ISO} allowed the first studies of the IR-galaxy LF 
at $z$$\lsimeq$0.3 (Saunders et al. 1990) and at $z$$\lsimeq$1, respectively (Pozzi et al. 2004). 
With {\em Spitzer} 24-$\mu$m data, it was possible to study the evolution of the mid-infrared (MIR) LF up to $z$$\sim$2.5 (e.g. Le Floc'h et al. 2005; Caputi et al. 2007; Rodighiero et al. 2010), while, even with the deepest {\em Spitzer} 70-$\mu$m data, only $z$$\sim$1 could be reached in the far-infrared (FIR) (Magnelli et al. 2009). 
Ground-based observations in the mm/sub-mm range, probing the evolution of the most distant ($z$$\gsimeq$2) and luminous dusty galaxies, have so far been limited to the identification of sources at the very bright end of the luminosity function (e.g., Chapman et al. 2005).
All of these works detected strong evolution in both luminosity and density, indicating that IR galaxies were more numerous and more luminous in the past. The derived rates of evolution of the IR sources are significantly higher than observed at any other wavelength. 
Since the rest-frame IR spectral energy distributions (SED) of star-forming galaxies and AGNs peak at 60--200 $\mu$m, to measure their bolometric luminosity and evolution with $z$ we need to observe in the FIR/sub-mm regime. However, the detection of large numbers of high-$z$ sources at the peak of their IR SED could not yet been achievable, due to source confusion and/or low detector sensitivity.\\
\indent In the present paper, we discuss observations performed with the PACS instrument (Poglitsch et al. 2010) onboard the {\em Herschel Space 
Observatory} (Pilbratt et al. 2010). {\em Herschel} is the first telescope allowing us to detect the FIR population to high redshifts ($z$$\sim$3-4) and to derive its rate of evolution by a detailed LF analysis. The PACS Evolutionary Probe (PEP, PI D. Lutz) Surveys are
designed specifically to this purpose. 
With the observations dedicated to the GOODS-N  by the PEP 
Science Demonstration Program (SDP), we first address the main goal of determining the evolution with 
redshift of the galaxy and AGN LF in the FIR domain. The availability of a deep multiwavelength catalogue in the GOODS-N is crucial for analysing the SEDs, obtaining k-corrections and total IR luminosities, and studying the LF and its evolution separately for the different IR populations. 
The measure of the total IR luminosity allows us to derive for the first time the total IR LF and its evolution directly from FIR data (see e.g. Nordon et al. (2010), this issue, for descriptions of the failure, at least at $z$$>$1.5, of previous total IR luminosity extrapolations from the MIR).\\
\indent Throughout this paper, we assume a $\Lambda$CDM cosmology with H$_0$\,=\,71~km~s$^{-1}$\,Mpc$^{-1}$, $\Omega_m$\,=\,0.27, and $\Omega_{\lambda}\,=\,0.73$. 

\section{Multiwavelength source characterisation}
\label{SecSED}
Our reference samples were the PEP SDP blind catalogues at 100 and 160 $\mu$m to the $3\sigma$ level (corresponding to $\sim$3 and 5.7 mJy, respectively) in the GOODS-N. The PEP sources were cross-correlated with the 24-$\mu$m and multi-wavelength catalogues by means of a likelihood ratio technique (Sutherland \& Saunders 1992).
We refer to the Appendix of Berta et al. (2010) for a detailed description of the data and the source identification. To maximise the fraction of identifications, we limited our study to the area covered by the ACS ($\sim$150 arcmin$^2$), obtaining 227 and 253 sources at 100 and 160 $\mu$m, respectively (216 and 237, respectively, with either spec- ($\sim$70\%) or photo-$z$).\\
\indent We used of all the available data to derive the SEDs of our PEP sources, which we interpreted and classified by performing a fit (using the {\em Le PHARE} software; Ilbert et al. 2006) with a semi-empirical template library (Polletta et al. 2007) representative of different classes of IR galaxies and AGNs (spirals, starbursts (SB), QSOs, Seyfert2/1.8, and composite AGN$+$SB objects).
Three main SED classes are found to reproduce most of our sources: spiral, SB, and AGN2 (Seyfert2/1.8). The spiral SEDs show no clear signs of enhanced SF or nuclear activity (see the online Fig.~\ref{Figonltempl}), the FIR bump being characterised by relatively cold dust (T$_{dust}$$\sim$20 K). On the other hand, SB templates are characterised by warmer (T$_{dust}$$\sim$40--45 K), more pronounced FIR bumps and significant UV extinction, indicative of intense SF activity. Templates of moderately luminous AGNs (representing Seyfert2/1.8 galaxies) are characterised by a ``flattening'' in the 3--10 $\mu$m spectrum (suggesting detection of an AGN in the wavelength range where the host galaxy SED has a minimum) and a FIR bump dominated by SF and intermediate (in terms of both energy and T$_{dust}$) between spirals and SBs. Although they can be considered as SF galaxies at the wavelengths relevant to this work, we prefer to refer to them as ``AGN2'' throughout the paper, to keep in mind that they probably contain an AGN, whose presence, though not dominant in the FIR, might be very important for analysis in other bands (i.e., in the X-rays or the MIR).
In general, the considered templates provide very good fits to the SEDs of our PEP sources. However, in $\sim$10--12\% of cases the observed SEDs are very well reproduced by the templates over the entire UV/optical/NIR/IRAC range, while they are systematically underestimated in the MIPS/PACS range. In these cases, the PEP 100- and 160-$\mu$m data in flux density are always higher by up to a factor of $\sim$4 than the template at the same $\lambda$'s. This happens mainly for the Seyfert2/1.8 templates (for about 40\% of the PEP sources fitted by these SEDs) and in less frequent cases also for the spiral ones. We therefore constructed three new templates with a rest-frame 0.1--15 $\mu$m spectrum similar to that of Polletta et al. (2007), but with a higher FIR bump, by averaging together (in $\lambda$-bins) the observed rest-frame SEDs (normalised to $K_s$ band) exhibiting an excess in the FIR and fitted by the same template (Seyfert2/1.8/Sdm; see an example in Fig.~\ref{Figonlnewt2} of the online edition). We also added to the library three new SB templates obtained by interpolating between the Sdm and the SB SED NGC6090 (to fill the large gap existing in the library between spirals and SBs). The final results of the SED-fitting are reported in the online Table~\ref{onlTabSED}.
Our SED-fitting results show a clear trend with $z$, with the sources becoming ``warmer'', ``more active'', and ``more IR luminous'' with increasing $z$ (see the online Fig.~\ref{Figonlratio}, where the 160-$\mu$m/100-$\mu$m flux ratio versus $z$ is plotted for the PEP 160-$\mu$m selected sources). In particular, spiral galaxies are found exclusively at low-$z$ ($<$1), dominating at $z$$<$0.4 and coexisting with AGN2 and SB galaxies at 0.4$<$$z$$<$1. The AGN2 are observed up to $z$$\sim$1.5--2, with a strong concentration at 0.4$<$$z$$<$1. The SB galaxies are found at any redshift in the range $z$$\sim$0.4--3.0 and dominate between $z$$\simeq$1 and $z$$\simeq$2.5. The composite objects and the QSOs are found exclusively at high-$z$ ($>$1.5 and $>$3, respectively). Although sample selection effects may also play a role in the observed trend with $z$, we conclude that sources with enhanced SF activity dominate at higher-$z$, gradually becoming ``colder'',``less active'', and ``less IR luminous'' as we move towards lower redshift.
Similar results were found by Gruppioni et al. (2008), who performed the same SED-fitting study presented here, but for a MIR-selected sample. The results of Gruppioni et al. (2008) were then used by Gruppioni \& Pozzi (2010, in preparation, hereafter GP2010) to construct a backward evolution model of IR sources, assuming separate evolutionary paths for the five populations described above.
   \begin{figure}
   \centering
   \includegraphics[width=8.8cm,height=4.5cm]{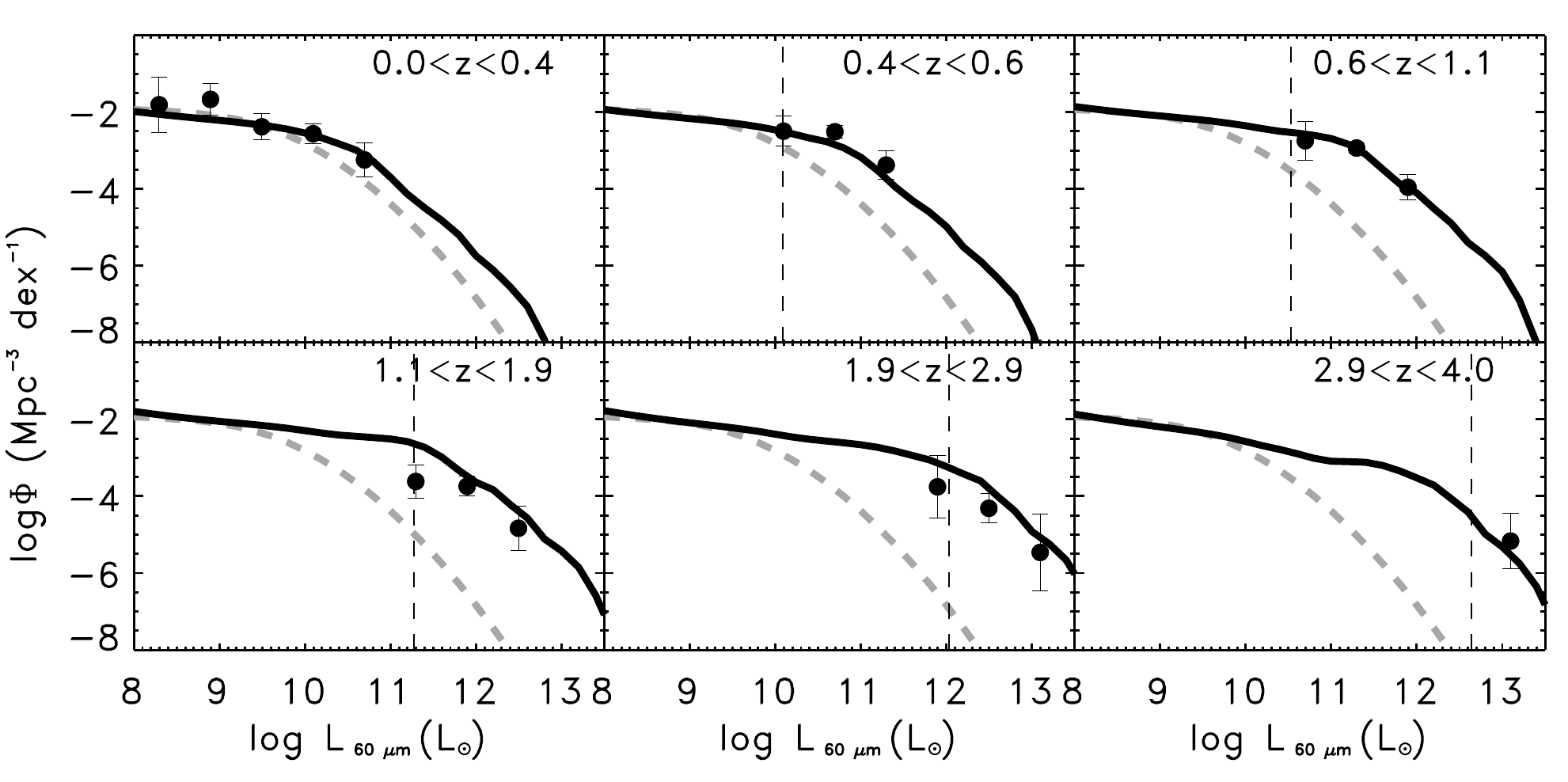} 
   \includegraphics[width=8.8cm,height=4.5cm]{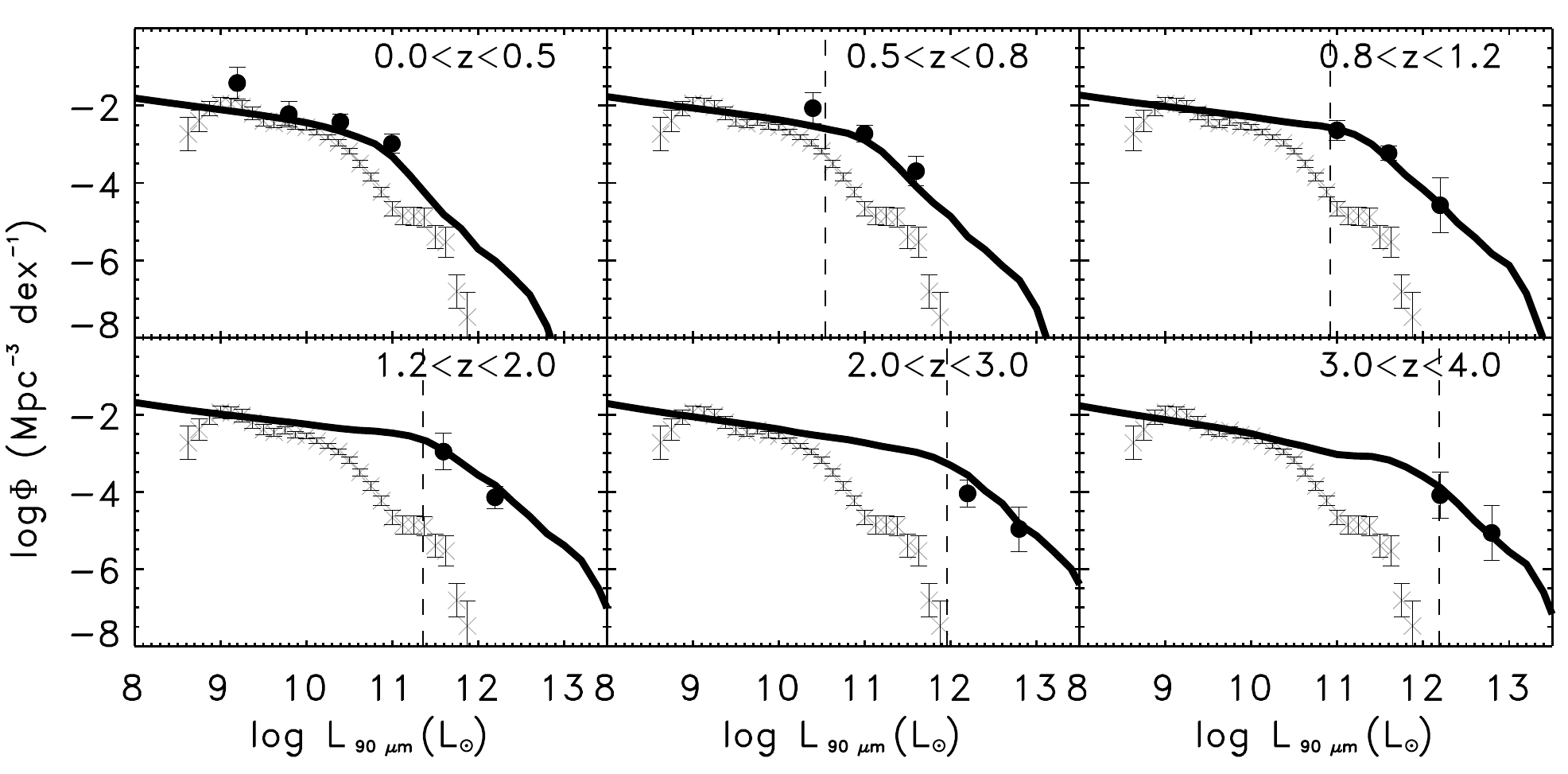} 
      \caption{Rest-frame LF at 60 $\mu$m ($top$) and 90 $\mu$m ($bottom$) in different redshift bins, computed as 1/V$_{max}$ (filled circles). The grey dashed line at 60 $\mu$m is the Saunders et al. (1990) LLF, while the diagonal crosses at 90 $\mu$m represent the LLF of Serjeant et al. (2004). The GP2010 evolutionary
           model predictions are also shown as a black solid line. }
         \label{FigLF}
   \end{figure}
\section{Luminosity functions}
\label{secLF}
\begin{figure}
   \centering
   \includegraphics[width=8.85cm,height=6.8cm]{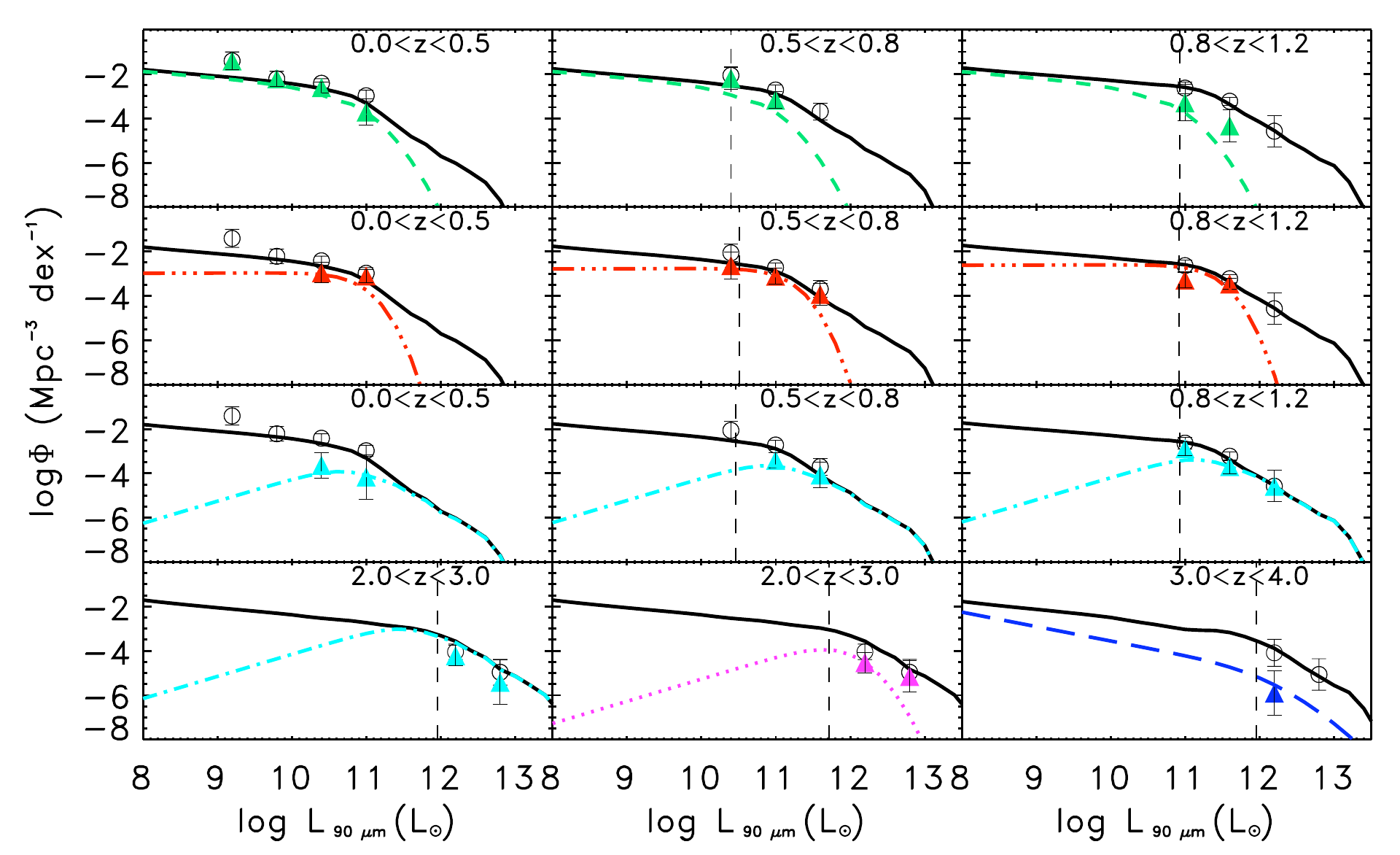} 
      \caption{Rest-frame 90-$\mu$m LF at different redshifts, 
      as computed from PEP (coloured filled triangles) and predicted by the GP2010 model for the different IR populations (green 
 dashed line: spirals ($three~top~panels$); red dot-dot-dot-dashed: AGN2 ($second~row$); cyan dot-dashed: SBs ($third~row~and~bottom~left~panel$); magenta dotted: composite ($bottom~middle$); blue long-dashed: AGN1 ($bottom~right$)). The total observed (open circles) and predicted (continuous line) LFs of Fig.~\ref{FigLF} ($bottom$) are reported. The vertical dashed lines are the luminosities above which we expect our survey to be complete in that $z$-bin for that population.}
         \label{FigLFpop}
   \end{figure}

The sizes of the PEP samples are sufficiently large to allow a direct determination of the FIR LF to $z$$\sim$2, and provide constraints on evolution to $z$$\sim$3--4. 
For the median redshift of the 100- and 160-$\mu$m PEP samples ($\sim$0.68 and 0.82 respectively), we computed the rest-frame LFs at 60 and 90 $\mu$m, using the $1/V_{max}$ method (Schmidt 1968). We divided the two samples into six redshift bins (slightly different for the two wavelengths) selected to be almost equally populated ($\sim$50 sources) at least up to $z$$\sim$2. In the two higher, less populated, redshift bins, the characteristic luminosity $L^{\star}$ is not constrained by data, but some hints of evolution can be detected (note that at $z$$>$3 most of the objects do have power-law SEDs and only photo-$z$'s, which are very uncertain due to the SED shape). 
The completeness and spurious correction factors given by Berta et al. (2010) 
were applied to our PEP data. The redshift incompleteness does not affect our conclusions, since $\sim$95\% of our sources have a redshift.
The results of the computation of our 60- and 90-$\mu$m LFs are shown in Fig.~\ref{FigLF} and reported in Tables \ref{TabLF100} and \ref{TabLF160} of the online edition of this paper.
The vertical dashed lines in each $z$-bin are the luminosities below which we expect our samples to be incomplete, given that at fainter luminosities not all galaxy types are observable (depending on their SED; Ilbert et al. 2004). 
For comparison, we overplot the LLFs at 60 $\mu$m and 90 $\mu$m from Saunders (1990) and Serjeant et al. (2004), respectively, and the predictions of the GP2010 backward evolution model. 
Strong luminosity evolution is necessary to explain the observed LF, even in the lower $z$-bin ($z$$<$0.4--0.5). 
To investigate the different evolutionary paths of the various IR populations, 
we computed the 1/V$_{max}$ LF separately for the five classes defined by the SED-fitting analysis. In Fig.~\ref{FigLFpop}, we show the 90-$\mu$m LFs derived from the 160-$\mu$m sample for the 
different SED classes, plotted in the most representative $z$-bins for each of them.  For comparison, we also show the single population LFs of the GP2010 model. The close agreement between data and model implies that both the evolution and the template SEDs (modified in the FIR as described in Sect.~\ref{SecSED}) assumed by the model (based on MIR results at $z$$\lsimeq$1.5) to represent the different IR classes can also be considered a fair representation of the FIR populations, at least up to $z$$\sim$3--4.  
In agreement with the change in population with $z$ implied by the SED-fitting analysis and discussed in Sect.~\ref{SecSED}, we observe that at lower redshifts ($z$$<$0.5) the LF is dominated by galaxies with spiral SEDs, while at 0.5$<$$z$$<$0.8 Seyfert2/1.8-SED sources appear to be more important than the normal galaxy population, becoming dominant around L$^\star$ (10$^{10.5}$L$_{\odot}$$\leq$L$_{90}$$\leq$$10^{11.5}$L$_{\odot}$) at 0.5$<$$z$$<$1.2. The SB population prevails at the higher luminosities (L$_{90}$$\geq$$10^{11.5}$L$_{\odot}$) at any redshifts, dominating also the L$^\star$ regime at $z$$\gsimeq$2. The composite and quasar populations are never dominant in the FIR, contributing to the 60- and 90-$\mu$m LFs mainly at $z$$>$2. 

\subsection{The total IR luminosity function}
\label{secLFtot}
We integrate the best-fit SED of each source at 8$\leq$$\lambda_{rest}$$\leq$1000 $\mu$m to derive the total IR luminosities (L$_{IR}$$=$L[8--1000 $\mu$m]). 
Our approach is similar to that of other studies based on MIR selected galaxy samples (i.e. Le Floc'h et al. 2005; Rodighiero et al. 2010), but this is the first time that the SEDs are accurately constrained by sufficiently deep data in the FIR domain and not simply extrapolated from the MIR to the FIR/sub-mm. 
%
   \begin{figure}
   \centering   \includegraphics[width=8.8cm,height=7.2cm]{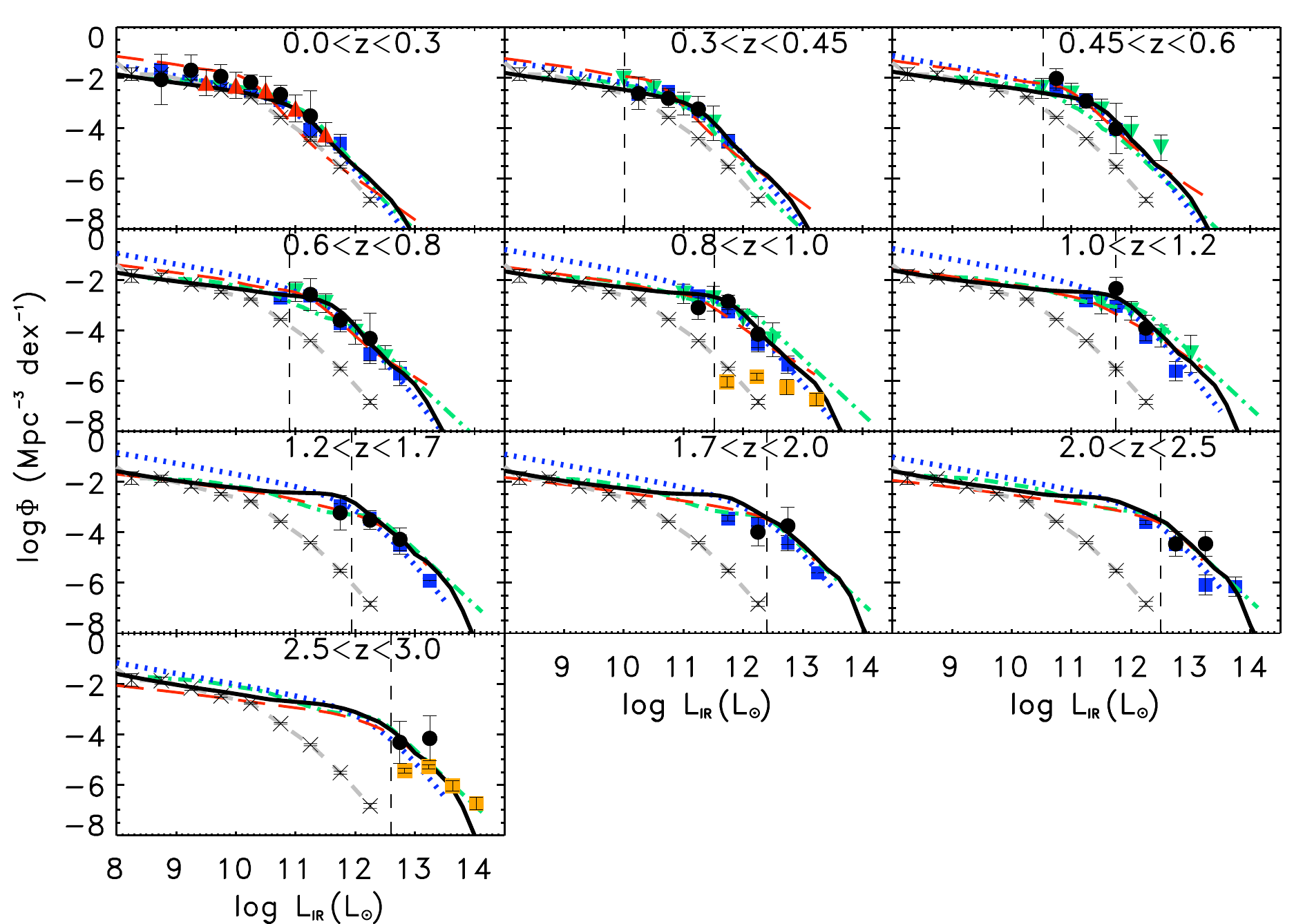} 

      \caption{Total IR LF from our PEP data (black filled circles). Other results from the literature are plotted for comparison (diagonal crosses: 
     LLF of Sanders et al. (2003); red filled triangles: Vaccari et al. (2010); green filled triangles: Le Floc'h et al. (2005); blue filled squares: Rodighiero et al. (2010); orange filled squares: Chapman et al. (2005)). The black solid, green dot-dashed, blue dotted, and red long-dashed lines represent the expectations from the GP2010, Lagache et al. (2004), Valiante et al. (2009), and Le Borgne et al. (2009) models, respectively.}
         \label{FigLFbol}
   \end{figure}
The total IR LF up to $z$$\sim$3 was computed (values reported in the online Table~\ref{TabLFtot}) and compared with other derivations available in the literature, as shown in Fig.~\ref{FigLFbol}. The total IR LF of Sanders et al. (2003) is plotted as a local reference, in addition to the predictions of four backward evolution models 
as references of evolution. Globally, data and model expectations seem to agree well over the whole $z$-range.
The agreement between the PEP total IR LF and those computed from 24-$\mu$m (ECDFS 0.3$\leq$$z$$<$1.2: Le Floc'h et al. 2005; VVDS+GOODS 0$\leq$$z$$<$2.5: Rodighiero et al. 2010) and SPIRE surveys (Hermes SDP: 0$\leq$$z$$<$0.2; Vaccari et al. 2010) is very good in the $z$-bins where the latter are available.  
The total IR LF of sub-mm galaxies from Chapman et al. (2005) at $z$$\sim$2.5 represents very well the bright end of the total IR LF, but seems to be affected by significant incompleteness at $z$$\sim$0.9. The sub-mm surveys clearly provide the most reliable probe of the evolution of the more distant and more luminous dusty galaxies, but are not sensitive enough to sample the low-intermediate redshift Universe.
Our PEP total IR LF evolves in luminosity as $\sim$$(1+z)^{4.1^{+0.3}_{-0.2}}$ up to $z$$=$1.5$\pm$0.3 (though considerable degeneracy is found between strict luminosity evolution and a combination of both density and luminosity). At 1.5$\lsimeq$$z$$\lsimeq$2.5--3, the evolution rate appears to remain almost constant. 
   \begin{figure}
   \centering
   \includegraphics[width=8.8cm]{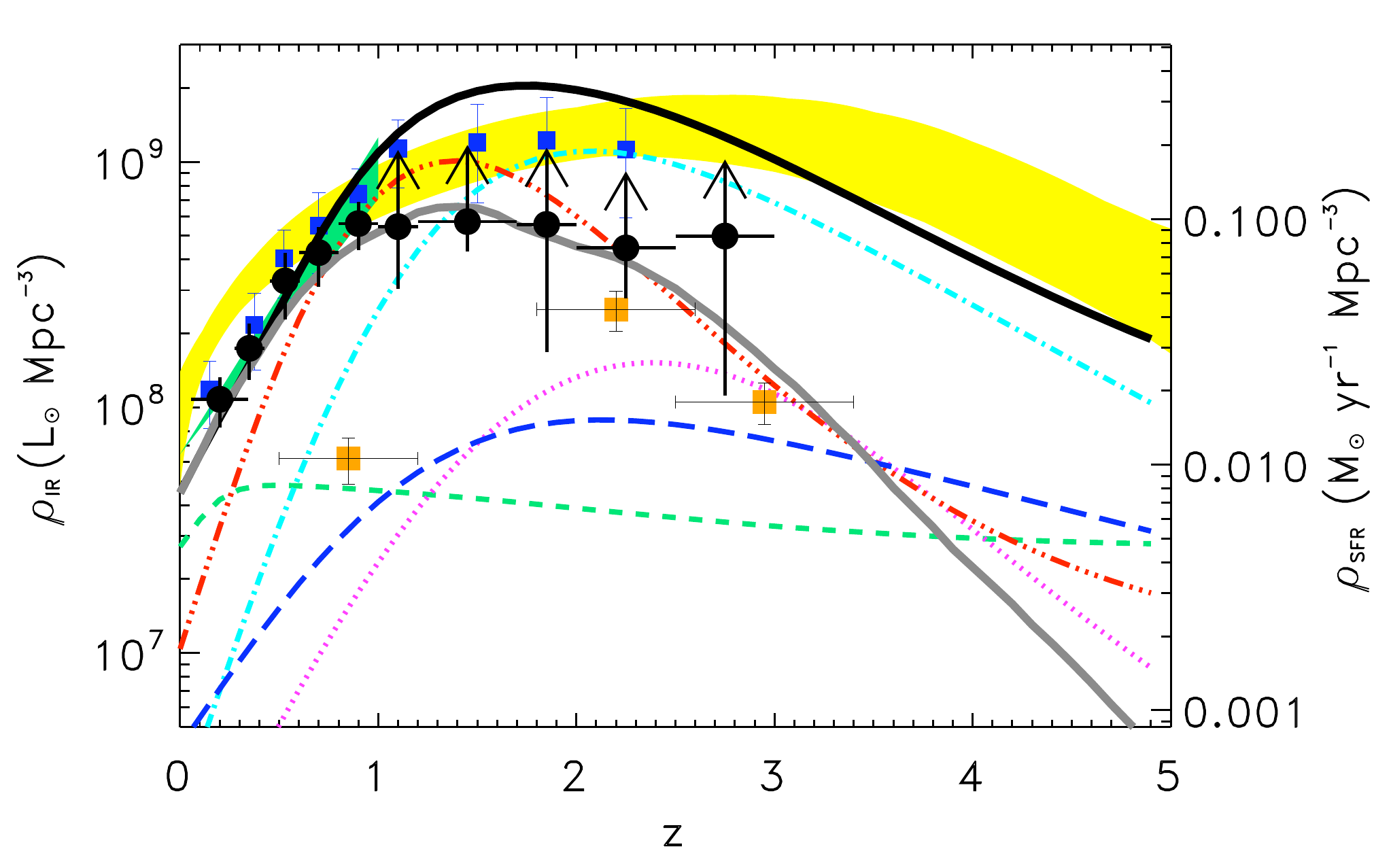} 
      \caption{Evolution of the total IR LD (or SFD) with $z$. The results of integrating our observed total IR LF in each $z$-bin are shown as black filled circles. 
      The 
      3$\sigma$ best-fitting envelope to the 
      Hopkins \& Beacom (2006) data collection from different surveys (yellow area), the results of Le Floc'h et al. (2005) and Rodighiero et al. (2010) as total IR LD up to $z$$\sim$1 (green area) and $z$$\sim$2.5 (blue filled squares), respectively, and the contribution to the SFD of radio detected sub-mm sources (orange filled squares) from Chapman et al. (2005) are also reported. The differently coloured lines are the GP2010 model predictions in terms of total IR LD (black solid), IR LD to the PEP luminosity completeness limit (grey solid), and single contributions from the different IR populations (same as in Fig.~\ref{FigLFpop}). }
         \label{FigLD}
   \end{figure}

\subsection{The comoving IR luminosity density}
\label{secLD}
One of the main reasons for computing the total IR LF is that it allows a direct estimate to be made of the total comoving IR luminosity density (LD) and the star formation density (SFD) as a function of $z$, which are crucial tools for understanding galaxy formation and evolution.
We convert L$_{IR}$ into star formation rate (SFR) by using the relation (for a Salpeter IMF) SFR [M$_{\odot}$ yr$^{-1}$] = 1.7$\times$10$^{-10}$ L$_{IR}$ [L$_{\odot}$] (Kennicutt, 1998), then
compute the SFD directly from our observed PEP LF, as well as from the GP2010 model.
Since the model provides a fair representation of the data, the predicted SFD can be interpreted as what we could expect from an ``ideal'' FIR survey (or collection of FIR surveys) covering the whole range of luminosities in each redshift bin. Our data are able to give a robust estimate of the LD only at $z$$\lsimeq$1--1.2, providing just lower limits at higher-$z$, given that with increasing redshift our total IR LF starts to become incomplete at the lower luminosities, the completeness limit moving towards the bright end of the luminosity distribution (at $z$$>$1, we sample just L$_{IR}$$\geq$L$_{IR}^{\star}$). The IR LD (or SFD) estimates from either data or model are shown in Fig.~\ref{FigLD}. To model the luminosity incompleteness of our data, we integrated the GP2010 total IR LF in each $z$-bin down to the completeness limit shown in Fig.~\ref{FigLFbol}, obtaining a SFD prediction (grey solid line of Fig.~\ref{FigLD}) that is totally consistent with our data estimates.
For comparison, we have also shown the SFD obtained from different surveys by Hopkins \& Beacom (2006) and the results of 24-$\mu$m surveys by Le Floc'h et al. (2005) and Rodighiero et al. (2010) and sub-mm surveys by Chapman et al. (2005). We find very close agreement with both previous results, based on either IR data or data at different wavelengths, and the GP2010 model predictions up to $z$$\sim$1. 
From PEP data, we find that the comoving IR LD evolves as $\sim$$(1+z)^{3.8\pm0.3}$ up to $z$$\sim$1. 
The model predicts a predominance of normal spiral galaxies only at very low redshifts ($z$$\lsimeq$0.3), when sources with Seyfert2/1.8 SEDs begin to dominate the SFD up to $z$$\sim$1.5. The SB galaxies then become the prevalent population up to the highest redshifts. We note that we have not subtracted the AGN populations from our SFD calculation, since the FIR part of the spectrum of all the templates 
(but the very few AGN1) is probably dominated by SF activity. Even in the Seyfert2/1.8 templates, the AGN activity dominates over the stellar emission only at $\lambda_{rest}$$\sim$3--10 $\mu$m. 
Given the power-law shape of their spectra, it is less clear at which $\lambda$ the AGN should dominate the light of the composite SEDs, and consequently 
what fraction of the total IR luminosity is related to accretion activity rather than SF in these objects. 
However, these populations are never predominant in our FIR survey and even a small fraction of contamination related to accretion activity occurring in these objects (mainly at high-$z$) will not significantly affect our results. 

\section{Conclusions}
We have used of the deepest {\em Herschel} 100- and 160-$\mu$m data from the PEP SDP survey in the GOODS-N to characterise the evolution of the galaxy and AGN FIR LF and LD across the redshift range 0$\lsimeq$$z$$\lsimeq$3. To summarise, in the present work we have:
   \begin{enumerate}
       \item completely characterised the multiwavelength SEDs of the PEP sources by performing a detailed SED-fitting analysis, 
       tightly constraining (and in some cases opportunely modifying) the templates at long wavelengths.
      \item computed the first rest-frame LFs at 60 and 90 $\mu$m up to $z$$\sim$4, finding that a significant amount of evolution is required to reproduce the LF at least up to $z$$\sim$2, the different SED-classes exhibiting different evolutionary behaviours.
      \item integrated the SEDs at $\lambda_{rest}$$=$8--1000 $\mu$m and computed the total IR LF up to $z$$\sim$3, 
      finding luminosity evolution $\propto$$(1+z)^{4.1^{+0.3}_{-0.2}}$ up to $z$$=$$1.5\pm0.3$, then an almost constant evolution rate to $z$$\sim$2.5--3.
      \item derived the evolution of the comoving total IR LD (and SFD), which was found to increase as $(1+z)^{3.8\pm0.3}$ up to $z$$\sim$1.
      Spiral galaxies are the main contributors to the SFD at $z$$\lsimeq$0.3, moderate SF galaxies (which most likely harbour a low luminosity AGN) becoming increasingly important up to $z$$\simeq$1.5. At this redshift, the SFD is representative of SB galaxies.
            \end{enumerate}

\begin{acknowledgements}
PACS has been developed by a consortium of institutes led by MPE (Germany) and including UVIE
(Austria); KU Leuven, CSL, IMEC (Belgium); CEA, LAM (France); MPIA (Germany); INAF-IFSI/
OAA/OAP/OAT, LENS, SISSA (Italy); IAC (Spain). This development has been supported by the
funding agencies BMVIT (Austria), ESA-PRODEX (Belgium), CEA/CNES (France), DLR (Germany),
ASI/INAF (Italy), and CICYT/MCYT (Spain). We thank an anonymous referee for a helpful report 
and L. Pozzetti and G. Zamorani for useful comments and for kindly reading the manuscript.
\end{acknowledgements}

%
\Online
\begin{appendix}
\section{Authors affiliations}\label{sect:affiliations}

\begin{enumerate}[label=$^{\arabic{*}}$]

     \item INAF--Osservatorio Astronomico di Bologna, Via Ranzani 1, I-40127 Bologna, Italy
     \item  Dipartimento di Astronomia, Universit\`a di Bologna, Via Ranzani 1, I-40127 Bologna, Italy
     \item   ESO, Karl-Schwarzschild-Strasse 2, D-85748, Garching, Germany
     \item INAF--Osservatorio Astronomico di Trieste, via Tiepolo 11, I-34143 Trieste, Italy
     \item  Dipartimento di Astronomia, Universit\`a di Padova, Vicolo dell'Osservatorio 3, I-35122 Padova, Italy
     \item ESAC, Villafranca del Castillo, ES-28692 Madrid, Spain
      \item CEA-Saclay, Service d'Astrophysique, F-91191 Gif-sur-Yvette, France
       \item Max-Planck-Institut f\"{u}r Extraterrestrische Physik (MPE), Postfach 1312, D-85741 Garching, Germany
      \item Instituto de Astrof{\'i}sica de Canarias, ES-38205, La Laguna, Spain
       \item Departamento de Astrof{\'i}sica, Universidad de La Laguna, Spain
        \item Department of Astronomy, 610 Space Sciences Building, Cornell University, Ithaca, NY 14853, USA 
        \item INAF--IFSI, Via Fosso del Cavaliere 100, I00133 Roma, Italy
        \item INAF--Osservatorio Astronomico di Roma, Via di Frascati 33, I--00040 Monte Porzio Catone, Italy
\end{enumerate}

\end{appendix}


\onltab{1}{
\begin{table*}
\caption{SED-fitting results}\label{onlTabSED}
\centering
\begin{tabular}{|c|c|c|c|c|c|}
\hline 
  $\lambda$ ($\mu$m)   &       Spiral         &   Starburst    & Composite  & AGN2           & AGN1 \\ \hline
 100                                 &     60 (28\%)   &   70 (32\%) &    13 (6\%)  &  71 (33\%)   &  2 (1\%)    \\
 160                                 &     63 (27\%)   &   82 (35\%)  &    13 (5\%)  &  76 (32\%)  &  3 (1\%)     \\
\hline
\end{tabular}
\end{table*}
}

\onltab{2}{
\begin{table*}
\caption{Rest-frame 60 $\mu$m luminosity function in the GOODS-N}\label{TabLF100}
\centering
\begin{tabular}{|r|cccccc|}
\hline  \hline
log$_{10}$L$_{60 \mu m}$ (L$_{\odot}$) & \multicolumn{6}{c}{log$_{10}$$\Phi$ (Mpc$^{-3}$ dex$^{-1}$)} \\
   & 0.0$\leq$$z$$<$0.4 & 0.4$\leq$$z$$<$0.6 &0.6$\leq$$z$$<$1.1 &1.1$\leq$$z$$<$1.9 &1.9$\leq$$z$$<$2.9 &2.9$\leq$$z$$<$4.2 \\ \hline
  8.3 &     $-2.03\pm0.73$     &                                     &                                     &                                    &                                      &      \\
  8.9 &   $-1.89\pm0.41 $     &                                     &                                     &                                    &                                      &      \\
  9.5 &    $-2.60\pm0.34$     &                                 &                                   &                                     &                                      &        \\
 10.1 &   $-2.79\pm0.25$    &  $-2.72\pm0.39$      &                                    &                                       &                                      &  \\
 10.7 &   $-3.47\pm0.45$    &  $-2.73\pm0.17$     &   $-2.97\pm0.50$      &                                        &                                      &         \\
 11.3 &                                  &   $-3.60\pm0.35$    &    $-3.16\pm0.14$     &   $-3.84\pm0.43$      &                                    &    \\
 11.9 &                                 &                                    &    $-4.18\pm0.33$     &   $-3.96\pm0.26$       &   $-3.98\pm0.81$     &  \\
 12.5 &                                 &                                    &                                    &  $-5.06\pm0.58$     &    $-4.54\pm0.38$  &  \\
 13.1 &                                 &                                    &                                    &                                      &   $-5.68\pm0.89$   & $-5.39\pm0.72$ \\
\hline\hline
\end{tabular}
\end{table*}
}

\onltab{3}{
\begin{table*}
\caption{Rest-frame 90 $\mu$m luminosity function in the GOODS-N}\label{TabLF160}
\centering
\begin{tabular}{|r|cccccc|}
\hline  \hline
log$_{10}$L$_{90 \mu m}$ (L$_{\odot}$) & \multicolumn{6}{c}{log$_{10}$$\Phi$ (Mpc$^{-3}$ dex$^{-1}$)} \\
   & 0.0$\leq$$z$$<$0.5 & 0.5$\leq$$z$$<$0.8 &0.8$\leq$$z$$<$1.2 &1.2$\leq$$z$$<$2.0 &2.0$\leq$$z$$<$3.0 &3.0$\leq$$z$$<$4.2 \\ \hline
 9.2   &  $-1.63\pm0.41$      &  &   &   &   & \\
 9.8   & $-2.44\pm0.33$     &  &    &   &     &       \\
10.4    & $-2.64\pm0.21$   &  $-2.28\pm0.40$      &                                   &                                &                                    & \\
11.0    & $-3.20\pm0.25$  &   $-2.95\pm0.21$     &    $-2.86\pm0.26$   &                                 &                                      &        \\
11.6     &                               &   $-3.92\pm0.38$     &   $-3.45\pm0.19$    &    $-3.18\pm0.47$ &                                 &                       \\
12.2      &                               &                                 &   $-4.80\pm0.71$     & $-4.37\pm0.29$   &   $-4.27\pm0.35$  &  $-4.31\pm0.61$ \\
12.8      &                                &                                &                                   &                                 &  $-5.19\pm0.58$   &  $-5.29\pm0.71$ \\
\hline \hline
\end{tabular}
\end{table*}
}

\onllongtab{4}{
\begin{landscape}
\begin{table*}
\centering
\caption{Total IR luminosity function in the GOODS-N}\label{TabLFtot}
\begin{tabular}{|c|cccccccccc|}
\hline  \hline
log$_{10}$L$_{IR}$ (L$_{\odot}$) & \multicolumn{10}{c}{log$_{10}$$\Phi$ (Mpc$^{-3}$ dex$^{-1}$)}  \\
         & 0.0$\leq$$z$$<$0.3 & 0.3$\leq$$z$$<$0.45 &0.45$\leq$$z$$<$0.6 &0.6$\leq$$z$$<$0.8 &0.8$\leq$$z$$<$1.0 &1.0$\leq$$z$$<$1.2 &1.2$\leq$$z$$<$1.7 &1.7$\leq$$z$$<$2.0 &2.0$\leq$$z$$<$2.5 & 2.5$\leq$$z$$<$3.0   \\ \hline
 8.75   & $-2.08$$\pm$0.98   &                              &                                  &                                 &                          &                             &                            &                             &                             &   \\
 9.25   & $-1.71$$\pm$0.61  &                                   &                                 &                                  &                         &                             &                             &                            &                             &    \\
9.75     & $-1.95$$\pm$0.47   &                                &                                 &                                 &                         &                             &                             &                            &                              &     \\
10.25     & $-2.18$$\pm$0.28  &  $-2.63$$\pm$0.64  &                                &                                 &                         &                             &                              &                            &                              &   \\
 10.75    & $-2.67$$\pm$0.38  & $-2.82$$\pm$0.38 &  $-2.04$$\pm$0.37   &                                &                           &                              &                             &                               &                              &   \\      
 11.25    & $-3.52$$\pm$0.98  &  $-3.25$$\pm$0.50 &  $-2.93$$\pm$0.29   & $-2.59$$\pm$0.63  &  $-3.11$$\pm$0.46 & $-2.91$$\pm$0.94  &                             &                              &                              &\\
 11.75   &                                &                               &   $-4.02$$\pm$0.98 &  $-3.61$$\pm$0.45  & $-2.86$$\pm$0.24 & $-2.36$$\pm$0.45 & $-3.23$$\pm$0.69 &                            &                              &   \\
 12.25   &                               &                               &                                &  $-4.33$$\pm$0.88 &  $-4.16$$\pm$0.71 & $-3.92$$\pm$0.50 &  $-3.52$$\pm$0.36 &  $-3.99$$\pm$0.56 &                              & \\
 12.75   &                               &                                &                               &                               &                             &                              & $-4.29$$\pm$0.45 &  $-3.75$$\pm$0.73  &  $-4.46$$\pm$0.50  & $-4.32$$\pm$0.83 \\
13.25     &                              &                                &                                &                              &                             &                              &                              &                               &   $-4.45$$\pm$0.50 &  $-4.18$$\pm$0.89 \\
\hline \hline
\end{tabular}
\end{table*}
\end{landscape}
}

\onlfig{5}{
\begin{figure*}
\centering
\includegraphics[width=11cm,height=6.4cm]{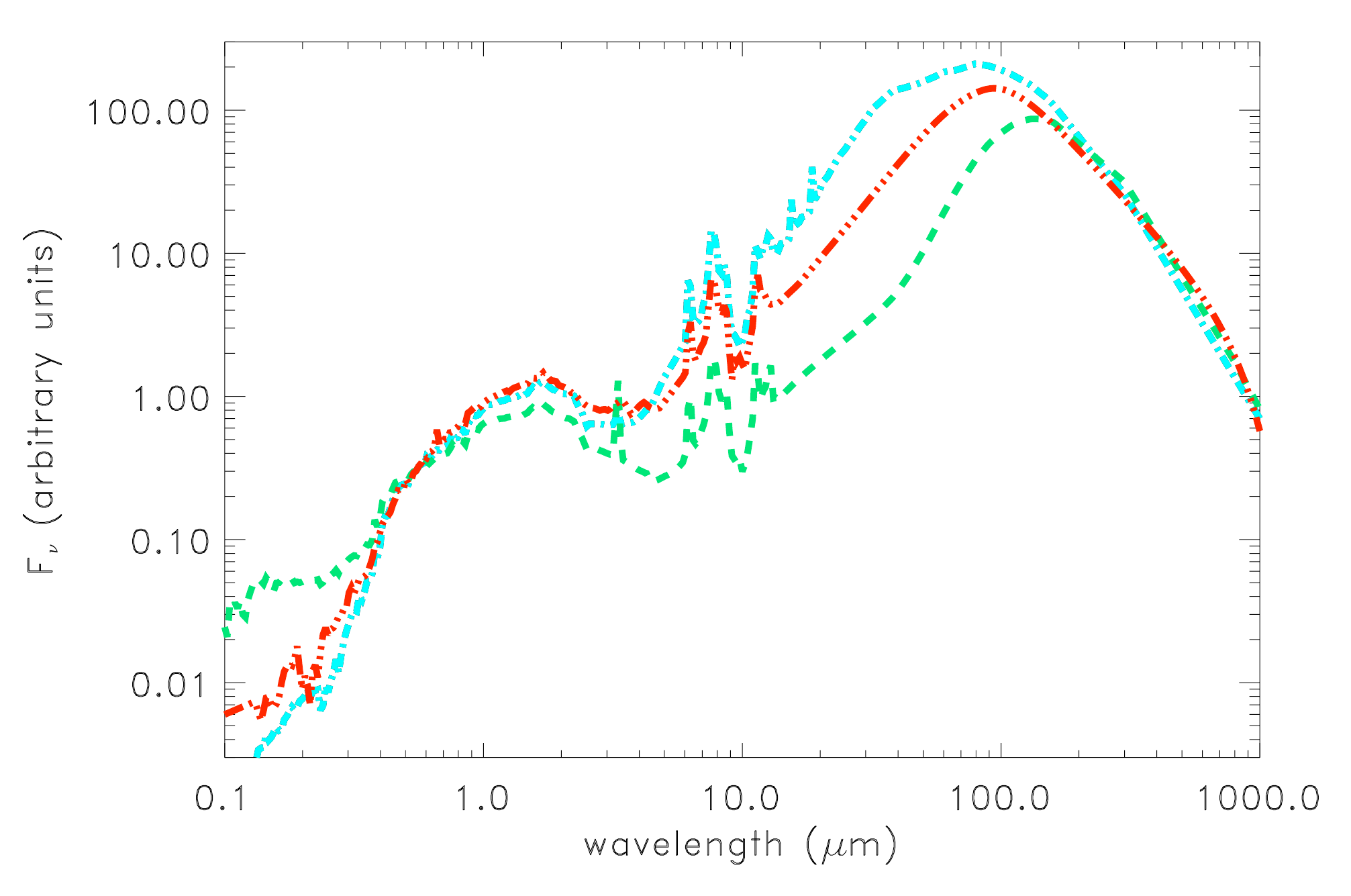} 
\caption{Template SEDs of a Starburst (M82: cyan dot-dashed line), a Seyfert 2 (red dot-dot-dot-dashed) and a Sdm galaxy (green dashed) from the Polletta et al. (2007) library, normalised to the $B$-band flux.}
\label{Figonltempl} 
\end{figure*}
}

\onlfig{6}{
\begin{figure*}
\centering
\includegraphics[width=11cm,height=6.5cm]{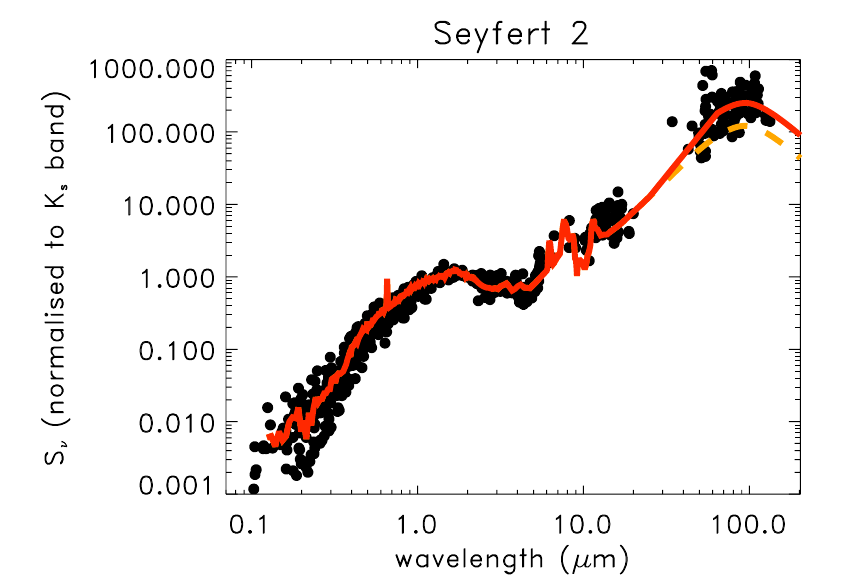} 
\caption{Rest-frame observed SEDs of the PEP 100- and 160-$\mu$m sources fitted by the Polletta et al. (2007) Seyfert 2 template (orange dashed line) with underestimated FIR. The red solid line shows the new ``FIR Seyfert 2'' template obtained by averaging the observed data in wavelength bins. }
\label{Figonlnewt2} 
\end{figure*}
}


\onlfig{7}{
\begin{figure*}
   \centering
   \includegraphics[width=11.cm,height=7.3cm]{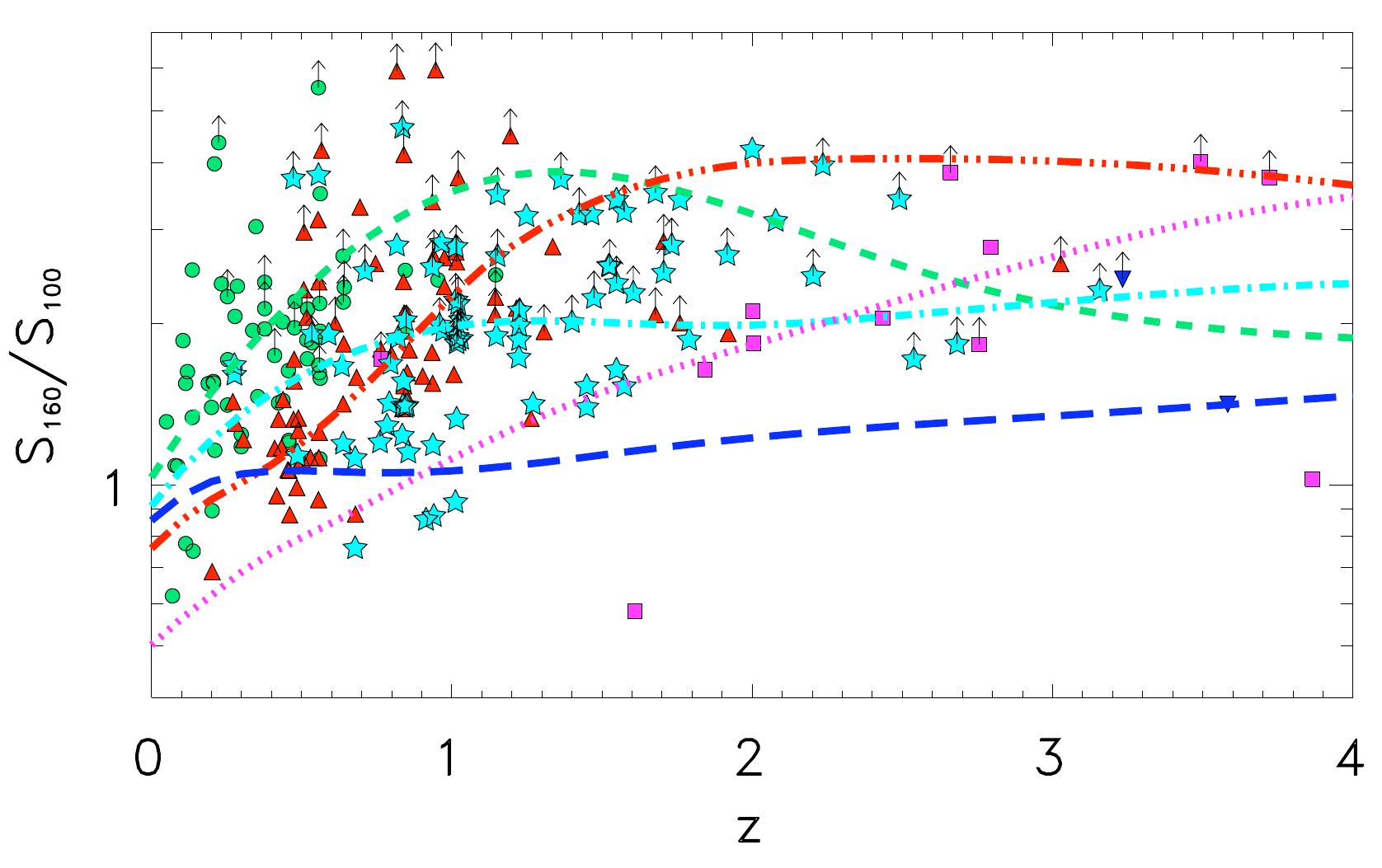} 
      \caption{160-$\mu$m/100-$\mu$m flux ratio versus $z$ both observed for the differently classified PEP 160-$\mu$m sources (symbols) and derived from the characteristic templates (curves): green circles and dashed line represent the normal galaxy population; cyan stars and dot-dashed line are for SB; red triangles and dot-dot-dot-dashed line are for AGN2; magenta squares and dotted line are for composite objects; blue triangles and long-dashed line are for AGN1. The upward arrows are for sources not detected at 100 $\mu$m (for which the 3 mJy flux limit has been considered).}
         \label{Figonlratio}
   \end{figure*}
}


\end{document}